\documentclass[fleqn,12pt,twoside]{article}
\usepackage[headings]{espcrc1}
\usepackage{graphicx}
\usepackage{epsfig}

\readRCS
$Id: espcrc1.tex,v 1.2 2004/02/24 11:22:11 spepping Exp $
\usepackage{graphicx}
\usepackage[figuresright]{rotating}

\newcommand{\AmS}{{\protect\the\textfont2
  A\kern-.1667em\lower.5ex\hbox{M}\kern-.125emS}}

\newcommand{\mueff}{\mu_{\rm eff}}
\newcommand{\beaa}{\begin{eqnarray*}}
\newcommand{\eeaa}{\end{eqnarray*}}
\newcommand{\bc}{\begin{center}}
\newcommand{\ec}{\end{center}}

\newcommand{\ep}{\epsilon}

\newcommand{\sg}{\sigma}

\newcommand{\Tr}{\mbox{Tr}}

\newcommand{\eela}[1]{\label{#1}\end{equation}}
\newcommand{\eeala}[1]{\label{#1}\end{eqnarray}}
\newcommand{\be}{\begin{equation}}
\newcommand{\ee}{\end{equation}}
\newcommand{\bea}{\begin{eqnarray}}
\newcommand{\eea}{\end{eqnarray}}
\title{Simulations of Cold Electroweak Baryogenesis\footnote{Talk
    presented at Strong \& Electroweak Matter (SEWM2006), BNL, United
    States, May 10-13, 2006.}}

\author{Anders Tranberg\address[DAMTP]{DAMTP, University of Cambridge, \\ 
        Wilberforce Road, Cambridge CB3 0WA, United
        Kingdom.}\thanks{Speaker. A.T. is supported by
        PPARC SPG ``Classical Lattice Field Theory''.},
 Jan Smit\address[ITFA]{Institute for Theoretical Physics, University of Amsterdam, \\ 
        Valckenierstraat 65, 1018XE, Amsterdam, The Netherlands.}
 and Mark Hindmarsh\address[SUSSEX]{Department of Physics and Astronomy, University of Sussex, \\ 
     Brighton BN1 9QH, United Kingdom.}
        }
       
\runtitle{}
\runauthor{}

\begin{document}

\maketitle

\begin{abstract}
We present real-time lattice simulations of Cold
Electroweak Baryogenesis, in which the baryon asymmetry of the Universe
is generated during tachyonic electroweak symmetry breaking at the end
of inflation. In the minimal realisation of the model, only three
parameters remain undetermined: the strength of CP-violation, the
Higgs mass and the speed of the symmetry breaking quench. The
dependence of the asymmetry on these parameters is studied.
\end{abstract}

\section{Cold Electroweak Baryogenesis}
Baryogenesis during a zero-temperature electroweak symmetry breaking
transition was proposed some time ago to bypass the lack of a first
order phase transition in the Standard Model electroweak sector
\cite{Garcia-Bellido:1999sv,Krauss:1999ng,Copeland:2001qw}, (see also
\cite{Turok:1990in,Turok:1990zg,Rajantie:2000nj,Copeland:2002ku,Garcia-Bellido:2002aj,Smit:2002yg,Skullerud:2003ki,Garcia-Bellido:2003wd,Tranberg:2003gi,vanTent:2004rc,vanderMeulen:2005sp,Tranberg:2006ip}
for related work).

Inflation is assumed to end near the electroweak scale ($100$ GeV), and
through coupling directly to the Higgs field, the inflaton is
responsible for triggering electroweak symmetry breaking. The ensuing
tachyonic instability supplies departure from equilibrium, which in the
presence of baryon number changing processes and CP-violation can lead
to a net baryon asymmetry. It is well known that non-perturbative dynamics of the electroweak
sector gauge fields may lead to baryon
number ($B$) violation through the anomaly equation (see for instance \cite{Rubakov:1996vz})
\be
\label{ncs}
\qquad B(t)-B(0)=3\langle \left[N_{\rm cs}(t)-N_{\rm
    cs}(0)\right]\rangle=\frac{3}{16\pi^2}\int_{0}^{t} dt\int
    d^{3}{\bf x} \langle\Tr F^{\mu\nu}\tilde{F}_{\mu\nu}\rangle, 
\ee
where the Chern-Simons number $N_{\rm cs}$ of the $SU(2)$ gauge field
is given in terms of the field strength $F^{\mu\nu}$ and its dual
$\tilde{F}^{\mu\nu}$. The CP-violation in the CKM matrix of the
    Standard Model is likely to be much too small to account for the observed asymmetry 
\cite{Gavela:1994ds,Gavela:1994dt}, and we shall here employ a
generic CP-violating term, possibly originating from some extension of
    the SM.

\subsubsection*{Model}

The simplest implementation of Cold Electroweak Baryogenesis is given
through the action
\be
S=
-\int d^{3}{\bf x}\,dt \bigg[\frac{1}{2g^{2}}\Tr F^{\mu\nu}F_{\mu\nu}+(D^{\mu}\phi)^{\dagger}D_{\mu}\phi
+\mu_{\rm eff}^{2}\phi^{\dagger}\phi+\lambda(\phi^{\dagger}\phi)^{2}+\kappa\,\phi^{\dagger}\phi
\Tr F^{\mu\nu}\tilde{F}_{\mu\nu}\bigg]\nonumber,
\ee
$\mueff$ is a time-dependent effective mass for the Higgs field. It
represents the Higgs coupling to the rolling inflaton, and we will
specialise to a linear form \cite{Copeland:2001qw,vanTent:2004rc},
%
%
$\mu^{2}_{\rm eff}=[\mu^{2}-\lambda_{\sigma\phi}\sigma^{2}]=\mu^{2}(1-2t/t_{Q})$
where $\sg$ is the inflaton
field, and we have introduced a quench time $t_{Q}$. $\delta_{\rm cp}$
will parametrise the strength of effective CP violation through
$\kappa=(3\delta_{\rm cp})/(16\pi^{2}m_{W}^{2})$, with $m_{W}$ the
W mass. The Higgs self-coupling fixes the Higgs mass through $\frac{m_{H}}{m_{W}}=\sqrt{8\lambda/g^{2}}$.

\section{Numerical results}
The numerical implementation follows exactly \cite{Tranberg:2003gi} and \cite{Tranberg:2006ip}. An ensemble of initial conditions is
evolved using the classical equations of motion, and the observables
averaged over the ensemble. These include the Higgs expectation value
$\langle\phi^{\dagger}\phi\rangle$, the Chern-Simons number
eq.~(\ref{ncs}) and the Higgs winding number,
\be
(i\tau_2\phi^*,\phi)=\rho U, ~~U\in SU(2),\\
N_{\rm w}=
\frac{1}{24\pi^2}
\int d^{3}{\bf x}
\, \ep_{ijk}
\Tr\bigg[(\partial_{i}U)U^{\dagger}(\partial_{j}U)U^{\dagger}(\partial_{k}U)U^{\dagger}\bigg].
\ee
$N_{\rm w}$ is the observable of choice since it is integer and equals the Chern-Simons number at low temperatures and
therefore late times. We shall assume
\be
\qquad \qquad \qquad B^{\rm final}-B(0)=3 \langle N^{\rm final}_{\rm
  w}-N_{\rm w}(0)\rangle.
\ee
\begin{figure}[h!]
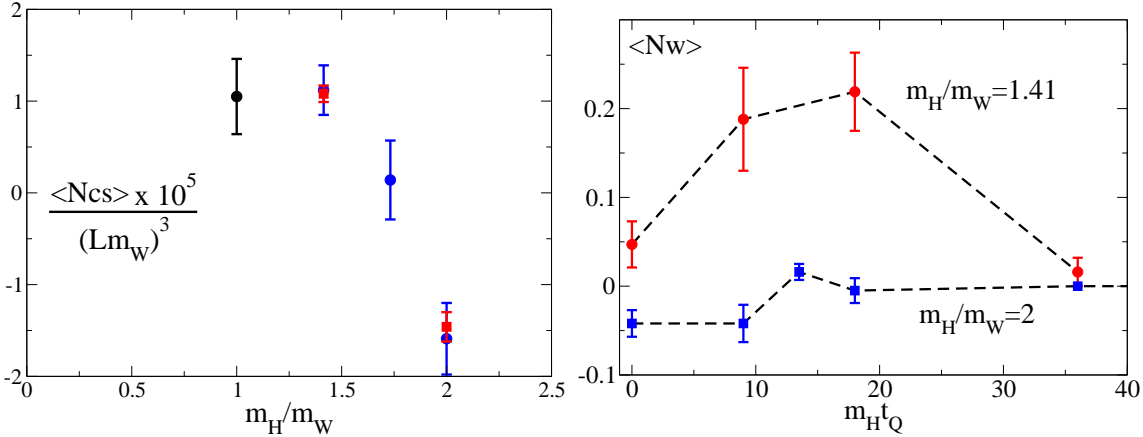

\epsfig{file=./pictures/3d_massdep_v2.eps,width=7.5cm,clip}
\epsfig{file=./pictures/resdivts24.eps,width=7.5cm,clip}
\caption{Left: Winding number density vs. $m_{H}/m_{W}$ for $t_{Q}=0$, $\delta_{\rm cp}=1$. The points
  with small
  error bars are results of linear fits to the $\delta_{\rm cp}$
  dependence (see figure~\ref{fig2}). Right: Higgs winding number $\langle N_{\rm w}\rangle$
  vs. $m_{H}t_{Q}$ for $\delta_{\rm cp}=1$, $m_{H}=2m_{W}$ and
  $m_{H}=\sqrt{2}m_{W}$.}
\label{fig1}
\end{figure}

\subsubsection*{Dependence on $m_{H}$}
In \cite{Smit:2002yg} and \cite{Tranberg:2003gi} we found that the mass dependence can be very
strong indeed. The overall sign of the final
asymmetry depends on it. Figure~\ref{fig1} (left) shows the final $\langle N_{\rm
  w}\rangle$ density at $\delta_{\rm cp}=1$, $t_{Q}=0$ for various Higgs
masses. Clearly the mass dependence is non-trivial. This turns out to be a
result of the conspiracy of the frequency of oscillations of the Higgs
and gauge fields. For a detailed study and modelling, we refer to
\cite{Tranberg:2006ip,Tranberg:2006??}. In the following, we shall consider
two specific values; $m_{H}=2m_{W}$, $m_{H}=\sqrt{2}m_{W}$.

\subsubsection*{Dependence on $t_{Q}$}
We can now vary the quench time $t_{Q}$ to find figure~\ref{fig1} (right). For
$m_{H}=2m_{W}$, there seems to be a fast quench regime
$m_{H}t_{Q}<10$, where the asymmetry is non-zero and roughly constant. For slower
quenches, the asymmetry becomes very small. Surprisingly, for
$m_{H}=\sqrt{2}m_{W}$ the intermediate range of quenches gives much
larger asymmetry (of the opposite sign) before decreasing to zero at
very slow quenches.

\subsubsection*{Dependence on $\delta_{\rm CP}$}
\begin{figure}
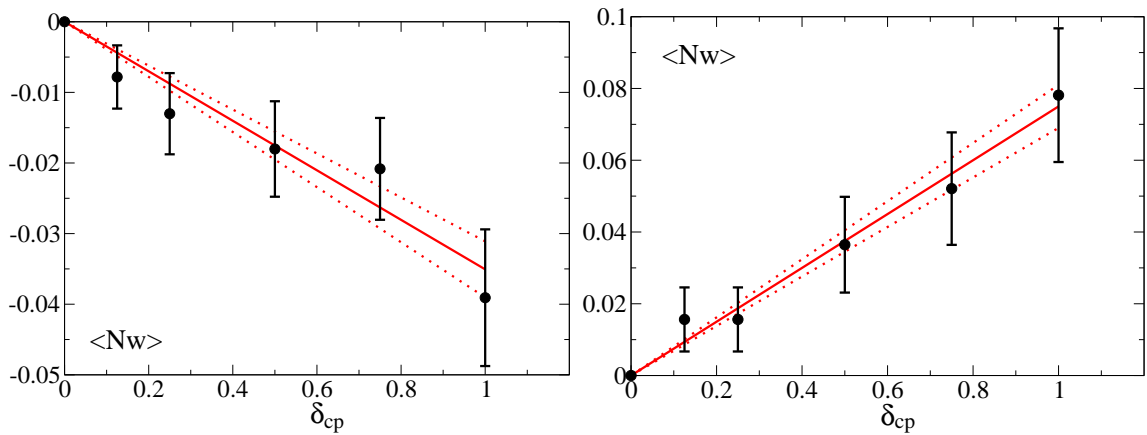

\epsfig{file=./pictures/kdep_s4_v2.eps,width=7.5cm,clip}
\epsfig{file=./pictures/kdep_s2_v2.eps,width=7.5cm,clip}
\caption{Higgs winding number $\langle N_{\rm w}\rangle$
  vs. $\delta_{\rm cp}$ for $t_{Q}=0$, $m_{H}=2m_{W}$ (left),
  $m_{H}=\sqrt{2}m_{W}$ (right). Full lines are
  best fits, dotted lines represent 1 $\sigma$ in the slope. This
  fitted slope gives the (red/square) points of figure~\ref{fig1}.}
\label{fig2}
\end{figure}

In figure~\ref{fig2} we show the dependence of $\langle N_{\rm w}\rangle_{\rm
  final}$ on the CP violation strength $\delta_{\rm cp}$ for $t_{Q}=0$
  and $m_{H}=2m_{W}$. In \cite{Tranberg:2003gi} we found that the dependence on
  $\delta_{\rm cp}$ was non-linear for large values of $\delta_{\rm
  cp}$ and we here present a zoom-in for smaller values, where the dependence
  is linear \cite{Tranberg:2006ip}. We will be interested in interpolating to very
  small values of $\delta_{\rm cp}$ when comparing to the observed baryon asymmetry.

\section{Conclusion}
In order to calculate the generated baryon asymmetry in terms of the
baryon ($n_{B}$) to photon ($n_{\gamma}$) density ratio, we use
\be
\qquad \qquad \qquad\frac{n_{B}}{n_{\gamma}} = 7.04\frac{n_{B}}{s},\quad s =
\frac{2\pi^2}{45}g_{*}T_{\rm
  reh}^{3},\quad\frac{\pi^{2}}{30}g_{*}T_{\rm reh}^{4}
=\frac{m_{H}^{4}}{16\lambda},
\ee
assuming that the final reheating temperature $T_{\rm reh}$ is given
by the initial energy density, distributed over the Standard Model
degrees of freedom, $g_{*}=86.25$ ($T_{\rm reh}<m_{W}$). With $Lm_{H}=27$ and putting
everything together, we for instance find for $n_{B}/n_{\gamma}$ at $t_{Q}=0$,
\bea
(0.40\pm0.03)\times 10^{-4}\times \delta_{\rm cp},~~~ m_{H}=\sqrt{2}m_{W},\quad
-(0.32\pm0.04)\times 10^{-4}\times \delta_{\rm cp},~~~ m_{H}=2m_{W}\nonumber.
\eea
The observed baryon asymmetry in the Universe 
$n_{B}/n_{\gamma}= 6.1\times 10^{-10}$ \cite{Spergel:2006hy} 
is reproduced for $\delta_{\rm cp}\simeq
2\times10^{-5}$ for $t_{Q}<10m_{H}^{-1}$ if the Higgs mass is around
$160$ GeV. In the lower end of the mass range, $m_{H}\simeq 120$ GeV, we can tune the
quench rate to increase the asymmetry by about a factor of two. Note
that this choice of Higgs mass also introduces an overall change of sign. This turns out be the result
of a conspiracy of oscillation frequencies and the complicated
dynamics associated with winding and un-winding of the Higgs
field. For detailed analysis of this behaviour, we refer to \cite {Tranberg:2006ip,Tranberg:2006??}.

\vspace{0.1cm}
\noindent{\bf Acknowledgements}: Part of this work was conducted on
the COSMOS supercomputer funded by HEFCE, PPARC and SGI. This work received support from FOM/NWO.


\end{document}